# Pinpointing Influential Posts and Authors in Social Media

Luiza Nacshon, Rami Puzis ,Amparo Sanmatu

**Abstract**

This research presents an analytical model that aims to pinpoint influential posts across a social web comprised of a corpus of posts. The model employs the Latent Dirichlet Allocation algorithm to associate posts with topics, and the TF-IDF metric to identify the key posts associated with each topic. The model was demonstrated in the domain of customer relationship by enabling careful monitoring of evolving "storms" created by individuals which tend to impact large audiences (either positively or negatively). Future research should be engaged in order to extend the scope of the corpus by including additional relevant publicly available sources.

## Introduction

Social influence has many areas of studies on how it can effect on the change in behavior that one person causes in another. The influence can be either positive or negative [1]. The social influence studies are important for businesses operations [2] or any other society functions [3]. In this paper we define influence as a diffusion discussion by highly referenced source post in social media.

Social media has become the most common tool for sharing thoughts, emotions, ideas, criticisms, etc. With social media on the rise each and every day, new discussion is published over the web, and an enormous number of users are exposed to it. The discussion can be published by individuals, companies, brands etc.

Some discussion is generated by ordinary users such are the posts in social media: blogs, online social networks, and internet forums. Other discussion is more official, coming from news websites and commercial pages. Some posts become highly shared or referenced, and end up drawing significant public attention and facilitate further discussions.

Companies are interested in identifying such discussions at early phases and monitoring them in order to align their campaigns and improve customer relationship and support branding efforts. These posts can then be studied for the purpose of obtaining business intelligence and analyzed for mining sentiments in order to estimate the public mood, etc.

When a company aims to identify popularity and importance of its posts on the social media, the company opens brand pages and follows its users' comments and likes on brand posts of that pages [4]. However following brand posts activities can't meet the needs of identifying public attention over the entire social media.

In a social media each entity can be influenced by others and vice versa, where entities can be users, pages, blogs etc. [5]. In addition, news webpages, blogs, tweets etc. can be influenced by most talked topics posted by others.

In this study we propose an analytical algorithm framework that pinpoints on key-posts which have a high influence on the online discussions.

The proposed method is suitable for any document-based dataset which contains the following four features: a unique identifier, content, a date of publication, author and references to other identified documents. For instance, in a dataset of posts from social media, the identifiers are URLs and the references are hyperlinks between pages.

After that the relevant fields are extracted from the given dataset, we employ standard topic detection algorithm on the content of the documents. The topic detection algorithm classifies documents which discussed same topics so called clusters. Note that one post can appear in many clusters.

The output of the proposed system is a set of influential documents in a cluster, in this paper the influential documents related to posts discussing a specific topic that are highly referenced in social media on that topic, referred as key-posts. Some of the selected key-posts may not even be included in the initial given dataset.

Finally, we analyze the timing of the reciprocal referencing among key-posts and visualize the discussion history of the elected key-posts, so called the network of key-posts. Note that the communities in the key-posts network identify discussion on a specific topic that was highly referenced. The graph of the key-posts helps to estimate the information diffusion in the social web in both topic and temporal dimensions, that is, highlighting, who is the source of the diffusion and who helped the diffusion by referencing the source.

We analyze a given dataset by applying a sliding window of days. The idea behind the sliding window is that we are able to elect more key-posts from the given dataset where each elected key-post was highly referenced in some small period of time.

Our goals in this research are, first, to analyze an existing corpus of forums, blogs, and micro-blogs by employing network analysis techniques in order to identify key-posts in each discussion topic. Next, we construct a social network graph that represents a webpages' network, where each node is a webpage's hyperlink and an edge connects between nodes that cite each other

The primary contribution of the current research is the application of standard Information Retrieval (IR) techniques in order to pinpoint the most influential topic-related documents (key-posts) for each one of the identified discussion topics. We employ the standard Term Frequency Inverse Document Frequency (TF-IDF) score [6] on webpages instead on set of words in order to select the key posts. The key-posts and the key-posts network emerging out of our analytic system will assists social media experts in tracking the diffusion of specific, high-profile discussions on the web. In addition, the key-posts content can be an input for sentiment analysis.

## Background

### Information Diffusion in social media

Social media provides platform for interactions among users in which they create, share, and exchange information and ideas in virtual communities. Such flows of information known as diffusion processes over social networks.

Information diffusion measures how information such as news, political abuse, complaints etc. propagates in social media. Such propagation mechanisms increase the likelihood that one user's activities (such as sharing content) can be amplified and influenced by others.

Social media analysis of information diffusion process i.e. aims to model how information propagates over a time sliding window, how fast and for how long, who are the key player among the users and which posts have an important role in the information diffusion process.

### Topic Detection in Social Media

In the literature, a topic is defined as a set of semantically related terms that express a single argument [11]. Topic detection refers to finding a group or community which discusses the same topic. In this study, we consider a diffusion of a discussion to be a large set of keywords (topics) that persists through a number of linked posts.

There have been several efforts in the area of community and topic detection in social media using term frequency [12] graph-based, [13] ,clustering using dictionary learning [14] and Latent Dirichlet Allocation (LDA) [15].

We focus on topic detection using LDA which is an algorithm for discovering topics that ocurr in some given contents. LDA represents set of documents as a mixture of contents (a.k.a bag of words) ,LDA then learns the topic representation of each document and the words associated to each topic.

In our work we apply LDA on a set of documents within a given dataset, where each document represents a post including extracted content. The output of LDA is a set of clusters, where each cluster represents posts discussing a similar topic.

### Information Retrieval in social media

Information Retrieval (IR) is method for obtaining information resources accurately as possible relevant to an information need from a corpus [7]. Nowadays there is an increasing interest of IR from social media. IR can be used in social media for example to identify and describe relations between webpages, the webpages content and semantic analysis.

There have been several efforts to apply IR models in the domain of social media analytics. For example using documents relevance score [8], or indexing text and semantic markup of webpages [9].

In this paper we use Term Frequency Inverse Document Frequency (TF-IDF) score for IR [10]. To model affinity of a postto a cluster which is a bunch of posts sharing the same distribution of words, we use TF-IDF score which was originally crafted to reflect the importance of a word with regard toto a document in a corpus. Term Frequency (TF) is the number of times a word appears in the document, while Inverse Document Frequency (IDF) is the number of times the word appears in the entire corpus.

In this study apply the TF-IDF score to pinpoint key-post by high affinity score. In our study TF – is the number of times a post was referenced by its cluster (), and IDF – how many times the same post was referenced by posts belonging to other clusters in the entire dataset. A high TF-IDF value leads to select the key-post i.e. select the post which is important across the cluster of posts, means that this posts was referenced by posts in specific cluster more than in other clusters and this defined as affinity of post to topic.

## Post to Topic Affinity

Let $R$ denote an input DataSet where each record $p \in R$ represents a post retrieved from the social media.

Each $p = (id_p, cont_p, d_p, auth_p, Refs_p)$ contains the following attributes:

* $id_p$ – URL that uniquely identifies $p$.

It is important to emphasize that multiple URLs can redirect to the same $p$. For instance URLs with query strings, fragments or shortened URLs, (e.g., http://en.wikipedia.org/w/index.php?title=TinyURL&diff=283621022&oldid=283308287 and its shortened version http://bit.ly/tinyurlwiki point to the same webpage [Wikpedia]). In this paper we ignore multiple pointers to the same $p$ by normalizing the URLs and following the redirections.

$cont_p$ - The main text of the post p (does not include comments, i.e. written statements that expresses opinions about post $p$).

$auth_p$ - The author of $p$.

$d_p$ - The time when p was published by its author.

$Refs_p$ - The set of hyperlinks within the main text of the post ($cont_p$). We also include in $Refs_p$ URLs that appear in the main text (e.g., alphanumeric sequences starting with http:// or https://). Note that the URLs in $Refs_p$ may be the identifiers $id_p$ of other posts in $R$.

Let $p_{first}$ and $p_{last}$ be the first and the last posts in $R$ respectively.
$d_{p_{first}} = min_{p \in R}\{d_p\}$
$d_{p_{last}} = max_{p \in R}\{d_p\}$
Let $|R|_d = d_{p_{last}} - d_{p_{first}}$ denote the time span of the posts in $R$.

We split the posts in $R$ into a series of overlapping sliding time-windows $W_1, \ldots, W_n$, each of size $\gamma$ with a step of $\delta$, where $n = \frac{|R|_d - \gamma}{\delta}$.

For each time window $W_i$, let $P_i$ denote the posts published during the time window:

$$P_i = \{p \in R : d_{p_{first}} + i\delta \leq d_p \leq d_{p_{first}} + i\delta + \gamma\}$$

Latent Dirichlet Allocation (LDA) is a standard topic detection algorithm that partitions a set of documents into clusters such that documents in the same cluster have similar word distributions. This algorithm requires predefined number of topics and the collection of documents (denoted by $k$). We apply $LDA$ for each $P_i$ throughout the dataset $R$. Let $T_i = LDA(P_i, k)$ denote the set of topics found by the LDA Topic Detection algorithm for a given $P_i$. Partition of the posts to topics may be crisp and non-overlapping meaning that each post belongs to one and only one topic. Some applications of LDA consider soft partitions where a post can belong to one or more topics with a degree of belonging ranging from zero to one. The proposed method can be used to find key posts in either case.

Next we define the post to topic affinity. According to Wikipedia the definition of affinity is a similarity of characteristics suggesting a relationship, especially a resemblance in structure between animals, plants, or languages. The general intuitive definition of posts to topic affinity is to meld group of posts who are knowledgeable about the same topics. In our study affinity quantifies the representativeness degree of a post to specific topic.
We use post cross-references to compute post affinity to topic by identifying influencers around particular topics of interest.
High affinity of a post p to a topic t means that p is frequently referenced within the topic t but is not referenced in other topics.

Let $T_{it} \in T_i$ denote a subset of posts in topic $t$ among all posts in a time window $W_i$.

URLs that appear as hyperlinks that are mentioned within the post will be represented as references inside the given dataset.
For the purpose of simplicity, we will continue referring to these URLs as posts although some of them may not be the URLs of posts within the input data set $R$.

We measure the affinity of these URLs to a topic using a measure similar to the Term Frequency Inverse Document Frequency (TF-IDF) score in information retrieval [10].

***URL Frequency*** – measures the extent to which a given URL ($url$) is cited in a given topic ($t_i$) of a given time window ($W_i$). Due to similarity to Term Frequency we will denote this measure as $TF(url, t_i) = |\{p \in T_{it_i} : url \in Refs_p\}|$

***Inverse Topic Frequency*** – measures the extent to which a URL ($url$) is cited throughout all topics ($T_i$).
Due to similarity to Inverse Document Frequency we will denote this measure as:
$IDF(url, T_i)$
$$= \log\left(\frac{|T_i|}{|\{t \in T_i \mid \exists (p \in T_{it} \wedge url \in Refs_p\}|}\right)$$

Given a URL ($url$) and a collection of topics ($T_i$) and a topic ($t \in T_i$) the affinity of $url$ in $t$ is defined as: $Repr(url, t, T_i) = TF(url, t) \cdot IDF(url, T_i)$.

Given a topic $t$ let $KP_i \subseteq P_i$ be the set of posts having the highest $Repr(id_p, t, T_i)$ value.
These posts affinity to the topic $t \in T_i$ since they are highly referenced (i.e. discussed) by the posts in $t$ but not by posts discussing other topics.
As a side effect of the proposed method we can identify key-posts that were not included in the original data set. Some of the posts appear as hyperlinks within the original post's main text, in the case that those hyperlink weren't extracted they will not be included in the data set. It is important to expand the data set for the data integrity of the elected key-posts and to find relevant posts despite keyword mismatch, publish date etc.

**Notations Table**

| Notation | Name | Meaning |
|---|---|---|
| $R$ | Data-Set | The input data set of social media pots |
| $P$ | Posts | Set of posts retrieved from the social media. Where each $p \in P$ Is an entry in $R$ |
| $id_p$ | URL | URL that uniquely identifies p |
| cont_p | Content | The main text of the post p |
| auth_p | Author | The author of p |
| d_p | Timeslot | The time when p was published by its author |
| Refs_p | Hyperlinks | The set of hyperlinks within the main text of the post (cont_p). posts that were referenced by p and are candidates to be elected as key-posts |
| T_i | Topics | Set of topics found by the LDA Topic Detection algorithm for a given P_i. Were i denotes a partition of posts from R. |

Let $G = (V, E)$ denote a key-posts Graph where $V$ – is the set of vertices and $E$ – is a set of directed unweighted edges.

Each key-post $url_{kp} \in Kp_i$ is a vertex V in G.
An edge E connects between two key-posts $url_{kp_1}$ and $url_{kp_2}$ if $url_{kp_1}$ referenced by $url_{kp_2}$. The size of V defines its $Repr(url, t, T_i)$ value.

For such a network G, its connected components can illustrate how data diffuses in the network, as they show which key-posts acknowledge one another discussing the same topic.

It is important to emphasize that $KP$ is a set of posts elected as a key-posts, that were not necessarily appeared in the initial data-set but were referenced by one or more $p \in R$.

### Key-Author Analysis

Key posts are published by key authors who are considered as topic representatives. The detection of key authors is done by calculating the aggregated TF-IDF score of each author's posts. **KP** is a given set of key-posts. For each $\mathbf{p} \in \mathbf{KP}$ we extract $\mathbf{Auth_p}$. For each $\mathbf{Auth_{kp}}$ we aggregate the **TF IDF** values such that:

$$Aggregated\_Influence(A) = (\sum_{p \in KP: A = Author_p} TF\,IDF_p)$$

Authors having highest TF IDF aggregated value for their key-posts are elected as key-authors.

## Author-Impact Analysis

The second author-based metric is the boost score of authors of posts $p \in P_i$. The boost score is important for the identification of negative wave incipience. The boost score determines whether authors cause the expansion of the wave they are taking part of. Therefore, our goal is to find the authors that by sharing a post create a buzz over this post.

Assume $auth_p$ referenced another post $p$ at time window $W_i$. $Auth_p$ is considered influential if post $p$ has been referenced by many other authors at the succeeding time windows.
To compute $auth_p$ impact on the popularity of a post $p$, we construct the array time slots_accumulated_counts. This array holds for each time window $W_i$ the number of distinct authors that referenced $p$ until and in time window $W_i$. It is important to emphasise that each $W_i$ represents number of distinct authors that referenced p until and inside this time window.

Computing boost score of $Auth_p$:

Per referenced post p we calculate the pointing score of $Auth_p$ on post $p$.

$$Score(ref) = \sum_{W_i > timeslot(ref)} \frac{accum(p, W_i) - accum(p, timeslot(ref))}{(W_i - timeslot(ref))^2}$$

$$AuthorBoostScore(Auth_p) = \sum_p \sum_{Auth_p} Score(ref)$$

accum (p, t ) – number of references to post p accumulated until time t.
*accum(p, timeslot(ref) –*

timeslot-?

The impact score of an $Auth_p$ is an aggregation over the impact scores of his or her references.

For example if $Bob_p$ is an opinion leader if the following scenario occurs consistently:
1. Bob references other user's post p on Facebook.
2. Post p becomes viral.

In other words, $Bob_p$ has high impact if he is the author of many boosting posts. A post q is considered a boosting post of a post p (that q references), if the publishing of q was followed by a boost in the referencing (re-posting) rate of p.

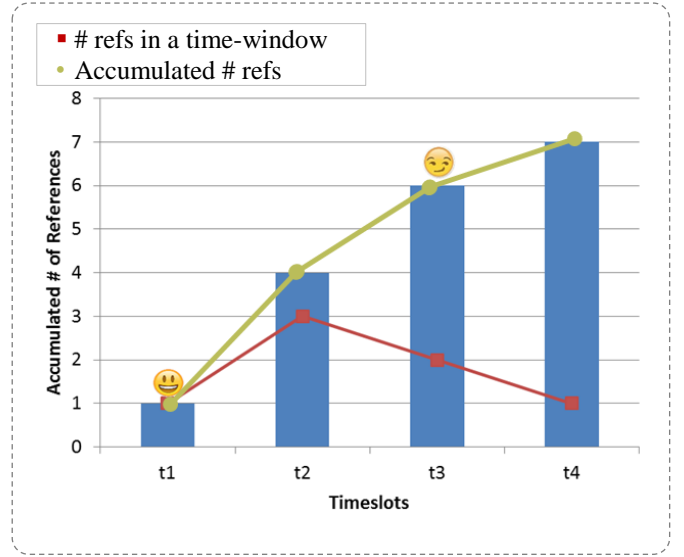

## System Overview

In this section we describe our system for electing key-posts discussing specific topics across a social media space. The pipeline of is the analytical system described in Figure 2. (1) For a given collection of post records, (2) divide the dataset by a selected sliding window into separate days and extract the necessary fields from the dataset (3) Cluster the posts based on specific discussion topics as generated by LDA (Fig. 3) (4) identify key-posts of each discussion using TF-IDF (Fig. 3) and generate the underlying network of topic-based communities and model the information diffusion (5) send posts that were missing in the initial data set to crawling system for obtaining the missing data (6) identify key-authors and generate ego-network which represents the connections between the key-authors and the authors point on key-authors (7) identify boost-authors and generate ego-network representing connections between boost-authors and the referenced authors.

The outputs of the analytical system algorithm include: (1) a set of key-posts having high TF-IDF score in a cluster where each cluster represents a discussion topic; (2) a set of key-authors have high aggregation over tf-idf over the elected key-posts; and (3) a set of boost-authors have high boost score.

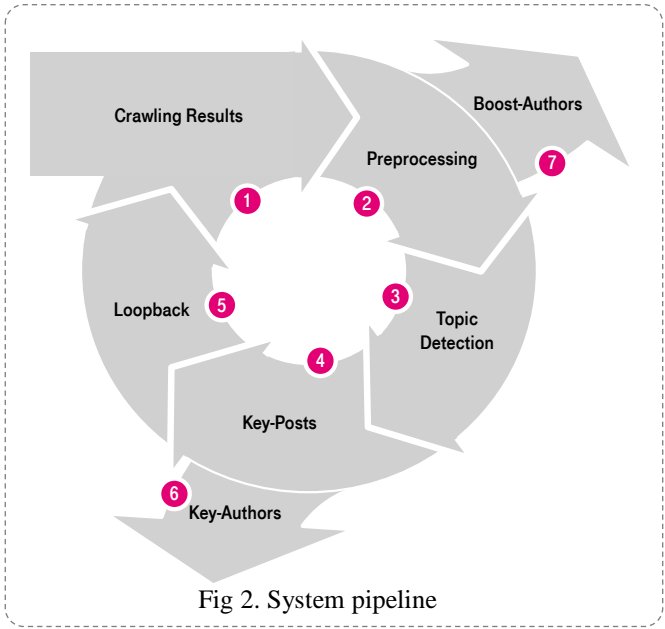

Fig 2. System pipeline

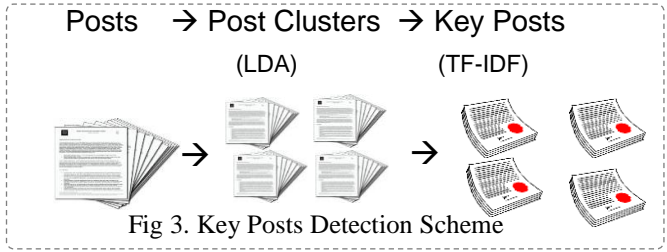

Fig 3. Key Posts Detection Scheme

**Data Preparation and Preprocessing**
The analytical system algorithm is suitable for handling structured data that follows the following schema where each record should include, a unique identifier; Content; Date of Publication; Author and References to other posts. The references can mention posts which are not part of the input dataset. We define the records in the input dataset as R where each record represents post p such that $p \in R$.

$$p = (id_p, d_p, auth_p, content_p, Refs_p).$$

Any input dataset should be preprocessed to fit the structured model while different domains can be adjusted to this model. For instance, in this paper we focus on data which consists of posts from online social media, the post's URL is the unique identifier; the post's text is the content; the date of publication is the date in which the post was made; the references are the hyperlinks appearing in the post possibly to websites which were not collected as part of the input dataset.

Preprocessing the data retrieved from the social web should include a link expansion stage that resolves the original URL from a shortened link. The expansion of the shortened URLs is critical for the accuracy of the key-posts election process. For example, a " http://bit.ly/tinyurlwiki " shortened link should be expanded to its final destination, so that several "bit.ly" links pointing to the same URL will not be considered as different URLs.

The second stage is to cleaning the content of the posts. The content retrieved from social media may contain a lot of unreadable chars, commas, tabs etc. and may contain a lot of extremely common words knows as stop words which should be removed from the text. The importance of stop words removal is in promoting accuracy of clustering posts discussing similar topics. If common words will be included in the detected topic many documents which are not really discussing the same topic will be clustered together.

The last stage of preprocessing is content stemming. The stemming is important for topic detection since many different posts may use different forms of a word but still discuss the same term. For example a stemming algorithm reduces the words "eating", "eater", and "eats" to the root word, "eat".

**Dividing the Dataset using a Predefined Sliding Window**
The public's attention evolves over time and discussion topics come and go. When analyzing a large dataset spanning over a long period of time, key-posts of discussions which did not last very long are considered to be weak and may not appear in the final results.

For instance, when analyzing data collected over a year, we may be interested in the records which received the most attention every week. For this purpose, the algorithm can be executed iteratively over subsets of posts selected using a sliding window.
Given a time window $w$, the data is split into batches consisting of W consecutive days. The algorithm is executed separately for each batch. At the end, a union of all the detected key-posts is used for creating the network of key-posts in time dimension.

**Topic Detection and Community Graph Construction**
Once the content was cleaned for each post $P_n$, we apply the LDA algorithm for clustering each post by its primary topic following several stages. (1) We first count the number of words in each post's content (2) the post and the number of words within its content are then forwarded as parameters to the LDA which generates a mapping of each post to a topic. (4) All posts associated with similar topics will be clustered together.

We create a undirected, weighted network (graph) where each post is a node. An edge connects two posts $p_1$ and $p_2$, if they appear in the same community and discuss the same topic. For each set of such posts, we collect their references. A reference may appear more than once in the network.

From this point, in which we have sets of references with their associated discussion topics, we proceed to the next stage of the algorithm which is election of the key-posts.

**Pinpointing Key Posts**

For each community's collection of referring posts, and for each reference, we calculate its TF-IDF score. We count the number of times it appears in topic (TF, URL frequency), and divide it by the total number of times it appears in all topics (IDF, inverse topic frequency). Thus, for each community we pick the key-post, the reference which received the highest TF-IDF score in the community's set of references.

Some relaxations can be made such as selecting the top references (i.e. more than one), however, the choice must be significant. For instance, if a set of references all receives the same (maximal) TF-IDF scores, none of them is more relevant than the others.

**Creating and Visualizing the Network of Key Posts**

It is usually beneficial to review whether key-posts are related to each other. For this purpose, the data for each of the key-posts may need to be collected if they were not part of the input dataset. A network can then be created where:
- Each key-post is a node,
- An edge connects two key-posts Kp1, Kp2 if Kp1 references Kp2.

For such a network, its connected components can illustrate how data diffuses in the network, as they show which key-posts acknowledge one another (since they reference each other). This requires obtaining, for each key-post, all of its properties described in the data model: it's content, date of publication and references. Visualization of this network, with the node sizes reflecting how many references were found for each record in the dataset ?and outside?, may help an analyst to review the findings and evaluate their correctness.

| Influence score | Measurement | Formula | Meaning |
| --- | --- | --- | --- |
| Post score | TF-IDF | $TF(url, t_i) =$ $\|\{p \in T_{it_i} :$ $url \in$ $Refs_p\}\|$ | Measures the post to topic affinity. Where TF- URL Frequency measures the extent to which a given URL is cited in a given topic. IDF - Inverse Topic Frequency measures the extent to which a URL is cited throughout all topics. |
| Affinity Score | Post to topic affinity | $Repr(url, t, T_i)$ $TF(url, t) \cdot$ $IDF(url, T_i)$ . | |
| Key Post | %x of max affinity of post to topic score | | Attracts a lot of attention in specific topic. We elect post as key post depends on the top x high affinity scored posts. Where x is a predefined number. |
| Key author | Max aggregation of affinity score of post to topic | | writes key posts |
| Boost score | | | The boost score determines whether authors cause the expansion of the wave they are taking part of. |
| Boost author | %x of max avg of authors boost score | | Shares posts that increase attention over short timeslot. |

**Table X.**

## Evaluation

In order to demonstrate the viability of our proposed social media analysis framework, we ran the system over a large corporate social media repository during March, 2015 and . Visualizing the relative importance of various key-posts in a given topic is demonstrated in Fig. 2. Each layer with a different color evolves over time (X) and its relative importance vs. other posts in the discussed topic (TF-IDF) is marked by its instant thickness (Y percentage out of 100). A different graph should be drawn for each topic and the graph should be constantly updated if the monitoring is done in real time (as opposed to retrospective batch summary for historical data).

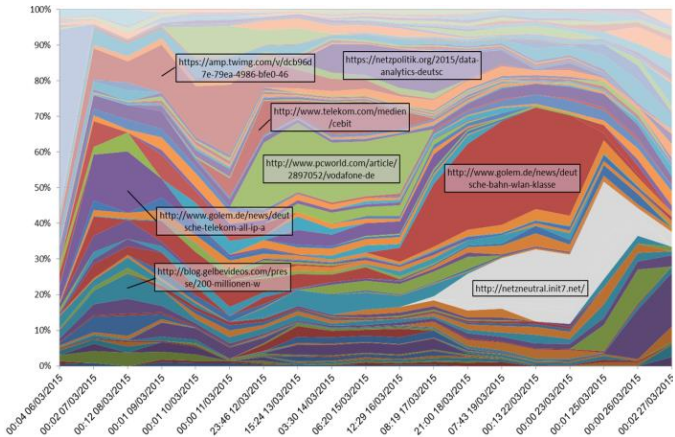

Fig 2. Key Posts of articles in March 2015

The result of the aggregated TF-IDF score for 20 key authors detected from the CRM system during March 2015 is presented in Fig. 3. The authors are classified using a color designating their relationship with the company (magenta, blue) and the nature of the post and its relevance to CRM (yellow, and framed yellow).

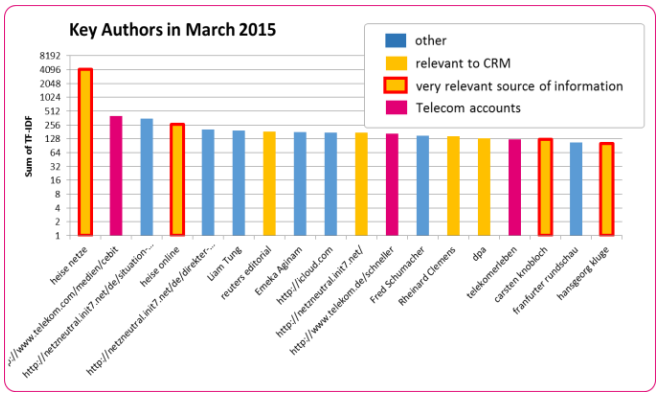

Fig 3: Key Authors March 2015

Fig. 4 depicts the average boost score of different types of authors denoted by the color. Our findings indicate that the top ranked boost authors refer to Twitter accounts. A lot of the highly ranked accounts belong to popular IT-News-sites, blogs or to the Telekom and all top ranked authors are highly active Twitter Users (> 1.000 #Tweets). The amount of followers vary among the top ranked authors.

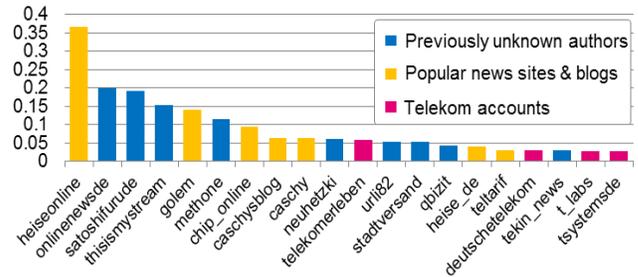

Fig 4: Boost Authors March 2015

Our findings demonstrate the viability of the diffusion tracking module based on the LDA as a dashboard instrument for monitoring storms in social media. Future research should be engaged in order to extend the scope of the corpus by including additional relevant publicly available sources, and identifying key-posts in general discourses. Metadata from the hyperlinked objects can be further used to improve classifier performance (Kinsella et al. 2011)

### Related Work more should be added

Ratkiewicz at el. (Ratkiewicz, Jacob, Michael Conover, Mark Meiss, Bruno Gonçalves, Alessandro Flammini, and Filippo Menczer, 2011) present Klatsch, a Machine Learning framework used to detect the early stages of viral spreading of political abuse. Klatsch uses a classification algorithm for the analysis of users' behavior. The system detects topics, represented by hash-tags, and builds a graph that represents the flow of information through Twitter. Using a classification algorithm the system can detect abusive political posts. Cataldi et al. (2010) also focus on twitter but rather extract the contents (set of terms) of the tweets and model the term life cycle according to a novel aging theory intended to mine the emerging ones. A term can be defined as emerging if it frequently occurs in the specified time interval and it was relatively rare in the past. Moreover, considering that the importance of a content also depends on its source, they analyze the social relationships in the network with the well-known Page Rank algorithm in order to determine the authority of posting users. Finally, they leverage a navigable topic graph which connects the emerging terms with other semantically related keywords, allowing the detection of the emerging topics, under user-specified time constraints.

Goyal et al. (Goyal A., Bonchi F., Lakshmanan L.V.S., 2008) present a data mining approach in social networks analysis for determining an influential user, termed a leader. Based on a frequency patterns, the algorithm discovers leaders, given the log of each users' actions on the social network and the social graph. The authors define an influence threshold and build an influence matrix to represent the users' influence on other users. The algorithm propagates through the social graph and for each node finds other nodes whose actions were influenced by current node. Similar to Goyal et al., our analysis identifies discussions (which they call "tribes") and key posts with different behavioral patterns ("leaders" and "tribe leaders"). However, our approach focuses on text analysis and is thus suitable for datasets in which the set of "actions" is very small or nonexistent, for instance, citation networks of scientific papersL'Huillier et al. (2010) address the topic-based community key members extraction problem, for which they combine both text mining and social network analysis techniques. This is examine how user popularity evolves. Each user is assigned a rank for each influence measure. Using the Spearman's rank correlation coefficient, the authors quantify how a user's rank changed across different measures and examine the high ranked users. While the authors we picked three of the most popular topics in twitter dataset 2009, we use the LDA algorithm for determine topics from tweets contents. Our goal as there is to find influence in social media. Both Meeyoung et al. (2010) and L'Huillier et al. (2010) search for influential users while our work focused on influential posts.

Vries et al. (De Vries, Lisette, Sonja Gensler, and Peter SH Leeflang, 2012 ) present method for investigating the factors affecting popularity of brand posts. The authors claim to be the first work on investigation of factors that influence the popularity of brand posts on social media sites. The factors are the visibility of the post, interactivity between the fans and the company, the content, location of the post and amount of likes and shares. The goal is to show how popularity of brand posts affect social marketing campaigns. Results indicate that the visibility, position and positive comments of the post affect the number of likes or shares, and influence on the popularity of the post. Similar to Vries et al. (2012), our general goal is to investigate influential posts in social media. While they aim to investigate the popularity and the influence of brand posts on marketing using a measure of post factor, we aim to investigate the evolution of topic-based influential posts across the social media in temporal-dimension. We measure the number of references each key post receives in some predefined time window to check how long it has been popular.

## Acknowledgments

The preparation of the files that implement these instructions was supported by The Live Oak Press, LLC, and AAAI Press.

Thank you for reading these instructions carefully. We look forward to receiving your electronic files!